\documentclass[english]{article}
\usepackage[latin9]{inputenc}
\usepackage[active]{srcltx}
\usepackage{textcomp}
\usepackage{amsmath}
\usepackage{graphicx}
\usepackage{wasysym}

\makeatletter

\DeclareFontEncoding{LGR}{}{}
\DeclareRobustCommand{\greektext}{%
  \fontencoding{LGR}\selectfont\def\encodingdefault{LGR}}
\DeclareRobustCommand{\textgreek}[1]{\leavevmode{\greektext #1}}
\ProvideTextCommand{\~}{LGR}[1]{\char126#1}

\newcommand{\lyxaddress}[1]{
	\par {\raggedright #1
	\vspace{1.4em}
	\noindent\par}
}

\makeatother

\usepackage{babel}
\begin{document}
\title{Supermassive black holes and dark halo from the Bose-condensed dark
matter}
\author{M. Morikawa and S. Takahashi }
\maketitle

\lyxaddress{Department of Physics Ochanomizu University 2-1-1 Otsuka, Bunkyo,
Tokyo 112-8610, Japan }
\begin{abstract}
Most of the galaxies harbor supermassive Black Holes (SMBH) in their
center. Some of them are observed in very high redshifts. We explore
the possibility that SMBH form from the coherent waves of Bose-Einstein
condensate (BEC) which are supposed to form the dark matter. We first
study the isotropic and anisotropic collapses of BEC. We find the
BEC wave can easily collapse to form SMBH but the realistic amount
of angular momentum completely prevents the collapse. We further explore
the Axion case with attractive interaction and find the moderate mass
ratio between the SMBH and the dark halo around it. We also obtain
the mass distribution function of BH within a single galaxy.
\end{abstract}

\section{Introduction}

Supermassive black holes (SMBH) of mass $10^{6-10}M_{\odot}$are
observed in the most of the galaxies\cite{kormendy2013}. All of them
are located in the galactic center and not a small amount of SMBHs
are already formed very early in the cosmic history of redshift $z\approx6-7.5$\cite{Banados 2018}.
Moreover, masses of the SMBH have strong correlation with the velocity
dispersion $\sigma$ of the galactic bulge $M_{BH}\propto\sigma^{4}$
\cite{kormendy2013}. All of these properties strongly indicate that
the SMBH plays the central role in the galaxy history and even defines
the center of the galaxy\cite{Morikawa 2015}. Then the principal
question would be the origin of these SMBH. Individual questions related
to the SMBH formation will be, 
\begin{enumerate}
\item (universality in the Universe) Why most of the galaxies harbor SMBH
of huge size$10^{6-10}M_{\odot}$?
\item (location in the galaxy) Why all the SMBH is located at the center
of the galaxy?
\item (causal relation with galaxy) Why SMBH form so early at least as $z\approx6-7.5$?
\item (correlation with galaxy) Why SMBH is firmly correlated with the galaxy
bulge $M_{BH\cdot}\propto\sigma^{4}$?
\item (correlation with dark halo (DH)) Why SMBH mass is correlated with
the galaxy dark halo mass $M_{DH}$ as $M_{BH\cdot}\approx10^{-4}M_{DH}$?
\end{enumerate}
There have been many literature in the past trying to answer these
questions. Most of them have been generally summarized in the diagram
in \cite{Rees 1984}. Basic mechanism in these is the gas collapse
to form the primordial black holes (BH), followed by the coalescence
of them or the accretion of the gas on them. Relatively heavy BH is
expected from the early population stars of mass $10^{1-2}M_{\astrosun}$
\cite{Volonteri2003}. The gas heating tends to prevent the effective
accretion. If a huge gas clump directly collapses to form a black
hole of size $10^{5-6}M_{\astrosun}$\cite{Oh2002}, then the Eddington
accretion rate can explain the early formation of SMBH although some
tuning of parameters is needed. 

All of these theories assume that the SMBHs are formed by baryons
or fermions. Contrary to them, we now would like to explore another
possibility that the SMBHs are formed by the dark matter (DM) or bosons,
in particular, by the quantum condensed boson fields such as the Bose-Einstein
condensation (BEC). The quantum condensation of bosons behaves very
differently from fermions, or from the fermion condensations, and
forms the macroscopic coherent wave. If the DM is formed from light
bosons, then it can easily form the quantum condensates, which may
coherently collapse to form SMBH in the early stage\cite{Nishiyama 2004}.
Then the stars and their clusters are formed around these SMBH afterward. 

All the above individual questions naturally suggest that the SMBH
defines the center of the galaxy first. SMBH can be formed before
stars and galaxies even at $z\approx10-20$, and the SMBH may trigger
the subsequent star and galaxy formation at later time $z\leq10$.
In other words, the co-evolution might be very rapid in the early
stage and the galaxy merger would not be the dominant mechanism at
the SMBH formation stage\cite{Heckman2014}. We now focus on the first
stage of this scenario, \textit{i.e.} the early formation of SMBH.
The subsequent star and galaxy evolution will be discussed in separate
articles, extending \cite{Morikawa 2015}, and including observational
predictions. 

Thus, we explore in this paper the possibility of the collapse of
the coherent boson field which may be the main component of the dark
matter (DM) \cite{Fukuyama 2006}. Our problem is now extended to
the question of how SMBH and dark halo (DH), being two different forms
of DM, are separated from each other? 

In the next section 2, we clarify how the BEC DM is possible and how
the condensation evolves in the Universe. In section 3, we explore
the collapsing dynamics in various conditions and show the BEC DM
actually collapses to form SMBH/DH. In section 4, we consider the
Axion model for BEC-DM and try to derive the time and mass scales
of the SMBH. In the last section 5, we conclude our study and describe
the subsequent scenario for the galaxy formation triggered by the
SMBH. 

\section{How do SMBHs form?}

 We now consider how the cosmic Bose-Einstein condensation (BEC)
is possible as dark matter (DM) and dark energy(DE) \cite{Fukuyama 2006,Fukuyama 2008,Fukuyama 2009,Schive2014}?
The critical temperature, below which the BEC takes place, is given
by 
\begin{equation}
kT_{cr}=\frac{{2\pi\hbar_{}^{2}n_{}^{2/3}}}{\zeta(3/2)^{2/3}m},\label{eq:Tcr}
\end{equation}
where $n$ is the number density of the boson particle of mass $m$
and $\zeta\left(3/2\right)\approx2.6$ is the zeta function. On the
other hand, the cosmic DM density evolution is given by 
\begin{equation}
\ensuremath{n=n_{0}^ {}\left({\frac{m}{{2\pi\hbar_{}^{2}}}\frac{T}{{T_{0}^ {}}}}\right)^{3/2}},\label{eq:cosmicDMT}
\end{equation}
where $T$ is the temperature of the DM and the suffix $0$ denotes
the present time. This expression for $n$ is obtained by the conservation
of the entropy $s$ per number density, 
\begin{equation}
\ensuremath{\frac{s}{n}=\ln\left({\frac{{e^{5/2}}}{{\left({2\pi\hbar^{2}}\right)^{3/2}}}\left({\frac{m}{T}}\right)^{3/2}\frac{{T^{3}}}{n}}\right)}.
\end{equation}
It is apparent that Eqs.(\ref{eq:Tcr},\ref{eq:cosmicDMT}) have the
same proportionality $T\propto n^{2/3}$.Therefore, once the Universe
enters into the phase of BEC, it stays in BEC in the later evolution
provided the process is adiabatic. The Universe would be mostly adiabatic
but locally violated, for example, in the violent process such as
the formation of SMBH by BEC. Therefore, for example if the boson
temperature was equal to the radiation temperature before the redshift
$\ensuremath{z=3000}$, $\ensuremath{T_{cr}^ {}=0.0027K,\rho_{0}=9.44\times10_{}^{-30}{\rm {g/cm}}_{}^{{\rm {3}}}}$,
then we would expect that the whole DM/DE is in BEC phase if $\ensuremath{m\leq10}$eV.
If the boson were the Axion field, then the temperature would be ultra-cold
and the BEC is inevitable.

BEC is described by the wave $\psi\left({t,{\bf {x}}}\right)$ which
evolves by the nonlinear Schroedinger equation\textit{ i.e.} Gross-Pitaevskii
equation (GP)\cite{Gross 1961,Pitaevskii 1961},

\begin{equation}
i\hbar\frac{{\partial\psi\left({t,{\bf {x}}}\right)}}{{\partial t}}=\left({-\frac{{\hbar^{2}}}{{2m}}\Delta+m\phi+g\left|\psi\right|^{2}}\right)\psi,\label{eq:GP}
\end{equation}
where $g=4\pi\hbar^{2}a_{s}/m$, $m$ is the boson mass, and $a_{s}$
is the scattering length, as well as the Poisson equation (PE)
\begin{equation}
\Delta\phi=4\pi Gm\left|\psi\right|^{2},\label{eq:PE}
\end{equation}
where $\phi$ is the gravitational potential. 

Both the above equations are the Newtonian approximation of the full
general relativistic formulation. However, as the first approximation,
this approach would be effective to identify the formation of BH by
the criterion that the amount of mass $M$ is compressed into the
corresponding Schwarzschild radius $r_{s}=2GM/c^{2}$. 

\section{SMBH from BEC}

The GP equation can be solved by the standard numerical methods \cite{Wimberger2005a,Wimberger2005b}.
However the simultaneous calculation of the parabolic GP and the elliptical
PE equation, is generally difficult to solve in particular for the
fast collapsing dynamics in which the solution may not converge to
any equilibrium finite form. Here we use semi-analytic calculations
for a general argument for the BH formation. Therefore we will further
make bold approximations. 

\subsection{Isotropic collapse}

The Lagrangian that yields the GP and PE is given by 

\begin{equation}
\begin{array}{cc}
L= & \left(\text{i\ensuremath{\hbar}}/2\right)(\psi^{\dagger}\dot{\psi}-\dot{\psi}^{\dagger}\psi)-\left(\text{\ensuremath{\hbar}}^{2}/2m\right)\nabla\psi^{\dagger}\nabla\psi-\left(g/2\right)\left(\psi^{\dagger}\psi\right)^{2}\\
 & -\left(1/8\pi G\right)\nabla\phi\nabla\phi-m\text{\ensuremath{\phi}}\psi^{\dagger}\psi.
\end{array}\label{eq:Lagrangian}
\end{equation}
In order to make the semi-analytic calculations possible, we use the
Gaussian approximation\cite{Gupta 2017}, 

\begin{equation}
\psi\left(t,x\right)=Ne^{-r^{2}/\left(2\sigma\left(t\right)\right)^{2}+ir^{2}\alpha\left(t\right)},\:\phi\left(t,x\right)=-\mu\left(\tau\right)e^{-r^{2}/\left(2\tau\left(t\right)\right)^{2}},\label{eq2}
\end{equation}
where $N$ is the number density of the boson particles. We roughly
estimate the BH formation when the portion of the DM enters inside
the corresponding Schwarzschild radius. 

Integrating this $L$ over the entire three-dimensional space, we
have the effective Lagrangian for the relevant variables $\sigma\left(t\right),\alpha\left(t\right),\mu\left(\tau\right)$,
and $\tau\left(t\right)$, 
\begin{equation}
\begin{aligned}L_{\mathrm{eff}} & =1/16(-(2\text{\textsurd2}gN^{2})/(\pi^{3/2}\sigma\left(t\right){}^{3})-(12N\text{\ensuremath{\hbar}}^{2})/(m\sigma\left(t\right)^{2})-(48N\text{\ensuremath{\hbar}}^{2}\alpha\left(t\right){}^{2}\sigma\left(t\right){}^{2})/m\\
 & +(32\text{\textsurd2}N\mu\left(t\right))/(\sigma\left(t\right)^{2}(2/\sigma\left(t\right){}^{2}+1/\tau\left(t\right){}^{2})^{3/2}))\\
 & -(3\text{\textsurd\textgreek{p}}\mu\left(t\right){}^{2}\tau\left(t\right))/G-24N\text{\ensuremath{\hbar} }\sigma\left(t\right){}^{2}\alpha'\left(t\right)).
\end{aligned}
\label{eq4}
\end{equation}
Then we can derive the ordinary differential equation, 
\begin{equation}
-\frac{\sqrt{2}gN}{\pi^{3/2}}+\frac{100}{81}\sqrt{\frac{10}{\pi}}Gm^{2}N\sigma(t)^{2}-6m\sigma(t)^{4}\sigma''(t)-\frac{6\hbar^{2}\sigma(t)}{m}=0,
\end{equation}
where the phase $\alpha\left(t\right)$ and the other variables are
dependent variables. It turns out, for the free case $g=0$, that
the effective potential $V_{\mathrm{eff}}(\sigma)$, derived from
the Lagrangian Eq.(\ref{eq4}), 
\begin{equation}
V_{\mathrm{eff}}(\sigma)=\frac{gN}{\sigma\left(t\right)^{3}}-\frac{Gm^{2}N}{\sigma\left(t\right)}+\frac{\hbar^{2}}{m\sigma\left(t\right)^{2}}
\end{equation}
has a minimum $\sigma_{min}$ beyond which the variable $\sigma\left(t\right)$
cannot reduce. The condition that the system of $\sigma\left(t\right)$
stays $\sigma_{min}$ forms a BH is $M>M_{kaup}$, where $M$ is the
total mass inside the radius of this minimum $\sigma_{min}$ and $M_{kaup}=0.633\hbar c/\left(Gm\right)$
is the Kaup mass. This mass is the critical mass at which the gravity
and the quantum pressure balance with each other. If this condition
holds, most of the mass turns into a black hole in the Gaussian approximation.
A typical numerical solution is given in Fig(\ref{fig1} Left). 

\subsection{anisotropic collapse}

\begin{figure}
\includegraphics[width=16cm]{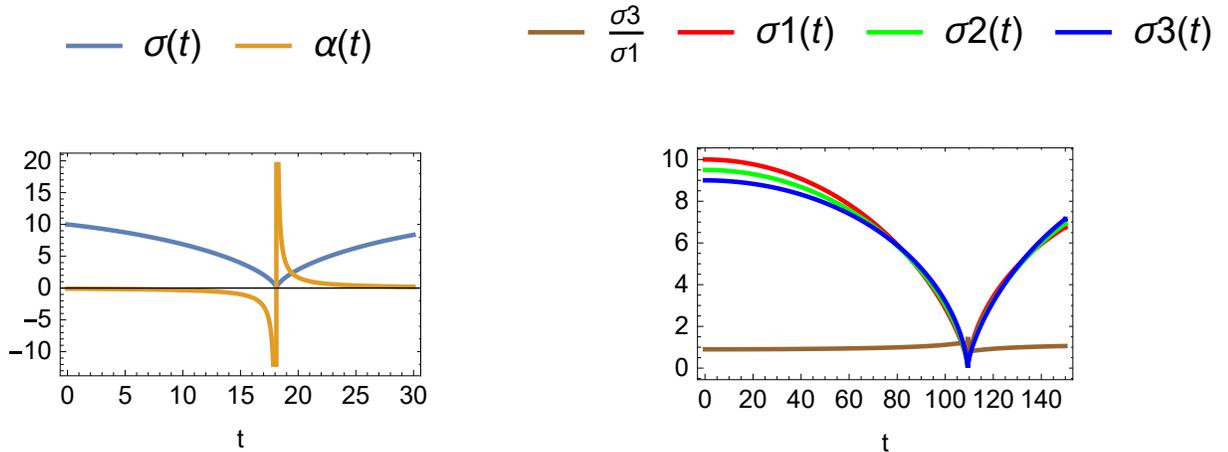}\caption{Time evolution of the typical collapsing dynamics of BEC. (left) Isotropic
collapse. The time evolution of the dispersion $\sigma\left(t\right)$
and the phase $\alpha\left(t\right)$of the condensate filed. If the
minimum size $\sigma_{min}$ is smaller than the corresponding Schwarzschild
radius, a BH is formed, otherwise it simply bounces. (Right) Anisotropic
collapse. The time evolution of the dispersion $\sigma_{i}\left(t\right)$
and the ratio $\sigma_{3}\left(t\right)/\sigma_{1}\left(t\right)$
of the condensate filed. Anisotropic BEC can easily collapse to form
BH as well. }
\label{fig1}
\end{figure}

It is easy to extend the above method to the anisotropic BEC collapse,
simply introducing independent dispersion for each spatial directions
$\sigma_{i}\left(t\right),\:i=1,2,3$, and the BEC field is expresses
as 
\begin{equation}
\psi\left(t,x\right)=\exp\left[i\mathrm{x_{1}^{2}}\alpha_{1}\left(t\right)-\frac{\mathrm{x_{1}^{2}}}{2\sigma_{1}\left(t\right)^{2}}+i\mathrm{x_{2}^{2}}\alpha_{2}\left(t\right)-\frac{\mathrm{x_{2}^{2}}}{2\sigma_{2}\left(t\right)^{2}}+i\mathrm{x_{3}^{2}}\alpha_{3}\left(t\right)-\frac{\mathrm{x_{3}^{2}}}{2\sigma_{3}\left(t\right)^{2}}\right].
\end{equation}
However, the effective Lagrangian becomes lengthy, 

\begin{equation}
L_{\mathrm{eff}}=-\frac{N}{4m\sigma_{1}(t)^{2}}-\frac{N\hbar^{2}}{4m\sigma_{2}(t)^{2}}-\frac{N\hbar^{2}}{4m\sigma_{3}(t)^{2}}-\frac{gN^{2}}{4\sqrt{2}\pi^{3/2}\sigma_{1}(t)\sigma_{2}(t)\sigma_{3}(t)}+\ldots-\frac{1}{4}mN\sigma_{3}(t)\sigma{}_{3}^{\mathrm{''}}(t),\label{eq:Lanisotropic}
\end{equation}
and the corresponding effective potential is ugly, 
\begin{equation}
V_{\mathrm{eff}}=\frac{gN}{2\sqrt{2}\pi^{3/2}\sigma_{1}(t)\sigma_{2}(t)\sigma_{3}(t)}+\frac{25\sqrt{\frac{10}{\pi}}GN\sqrt[3]{\sigma_{2}(t)}\sqrt[3]{\sigma_{3}(t)}}{243\sigma_{1}(t)^{5/3}}+\ldots+\frac{\hbar^{2}}{2m\sigma_{1}(t)^{2}}+\frac{\hbar^{2}}{2m\sigma_{2}(t)^{2}}+\frac{\hbar^{2}}{2m\sigma_{3}(t)^{2}},
\end{equation}
though being still useful for a rough analytic estimates and numerical
calculations. A typical numerical solution is given in Fig(\ref{fig1},Right).
Even in the anisotropic case, BEC can collapse to form SMBH. Unfortunately,
at the first collapse, most of the DM turns into SMBH and almost no
DH is left behind if we neglect the dissipation or angular momentum. 

\begin{figure}
\includegraphics[width=14cm]{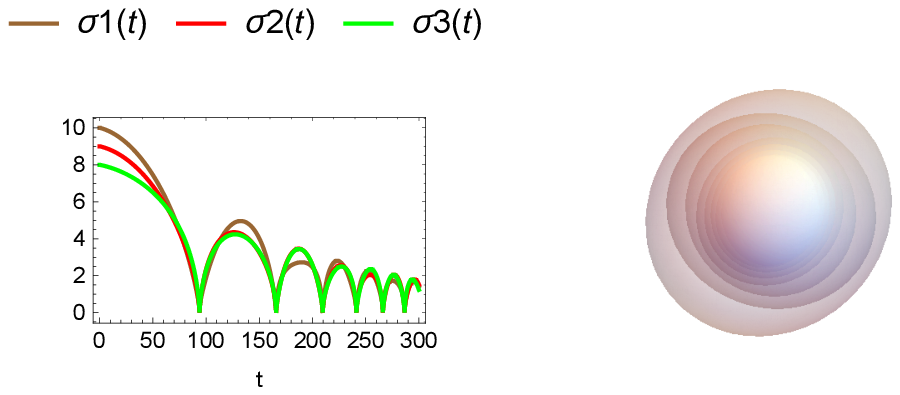}\caption{Time evolution of a dissipative anisotropic BEC collapse. (Left) A
damping collapse of BEC. This is a typical solution derived from the
effective Lagrangian Eq.(\ref{eq:Ldiss}). (Right) The superposition
of the snapshots of BEC field at each maximum expansion. It forms
the concentric almost spherical shells around the formed SMBH. }
\label{fig2}
\end{figure}

Further, it would be more realistic to consider the dissipative collapse.
Usually, the decay of the BEC into the normal gas is expressed by
an extra term $\gamma\psi\left(t,x\right)$ on the left-hand side
of Eq.(\ref{eq:GP}), where $\gamma$ is a constant. However, such
dissipation cannot be expressed in the proper Lagrangian. We now introduce
an explicitly time-dependent factor $e^{\gamma t}$ as 
\begin{equation}
L_{\mathrm{diss}}=e^{\gamma t}L\label{eq:Ldiss}
\end{equation}
where $L$ is the original Lagrangian Eq.(\ref{eq:Lagrangian}). Integrating
this $L_{\mathrm{diss}}$ over the entire space as before, we have
the effective Lagrangian which yields the dissipative equation of
motion. A typical numerical calculation of this is in Fig(\ref{fig2}
Left). 

Superposition of the snapshots of BEC field at each maximum expansion
is given in Fig(\ref{fig2} Right). A set of concentric shell structure
is conspicuous. The shell may form a temporal potential well, which
induce the active star formation there. If this structure is supported,
for example, by the rotation of the gas, this may leave some trace
such as the concentric shell structure of stars in the later evolution
of the galaxy\cite{Ebrov=0000B4a 2013}. 

\subsection{collapse with angular momentum}

The angular momentum of the system generally has a tendency to prevent
the BEC collapse into SMBH. However, an appropriate amount of angular
momentum may be important to get the coexistence of SMBH and DH, because
the whole system would collapse into BH otherwise. In this context,
from various observations, the individual angular momentum $A$ of
the cosmological objects shows the following scaling \cite{Nakamichi 2010},
\begin{equation}
A=\kappa\frac{G}{c}M^{2},\quad\kappa\approx10^{4}.\label{eq:Aobs}
\end{equation}
 In our case, if we start from the wave function 
\begin{equation}
\psi(t,\mathbf{x})=e^{-\frac{r^{2}}{2\sigma(t)^{2}}+ir^{2}\alpha(t)}Y_{l}^{m}(\theta,\phi),
\end{equation}
the effective potential becomes 
\begin{equation}
\frac{gN}{\sigma\left(t\right)^{3}}-\frac{Gm^{2}N}{\sigma\left(t\right)}+\frac{A^{2}}{m\sigma\left(t\right)^{2}}.\label{eq:VeffgGA}
\end{equation}
Therefore, if $g=0$, the minimum of this potential yields the maximally
allowed angular momentum for the BH formation. It turns out to be
$A_{\mathrm{max}}=2\frac{G}{c}M^{2}$ which is, unfortunately, well
below the observed value Eq. (\ref{eq:Aobs}) and therefore no BH
is formed at all. 

From this situation, we need to consider much dominant interaction
term with $g<0$, the attractive force which overcomes the prevention
of the collapse to SMBH by the angular momentum. There is another
possibility that the SMBH is formed in much earlier stage when the
over-density region acquires its angular momentum through the tidal
torque mechanism\cite{Sugerman 2000}. This latter possibility will
be studied in a separate article. 

\section{Collapse of Axion field -attractive interaction-}

\begin{figure}
\includegraphics[width=14cm]{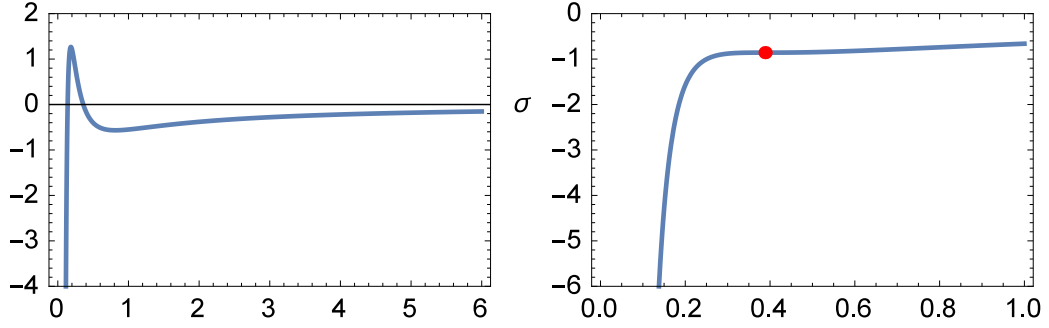}\caption{The effective potentials for the Axion DM. (Left) There is a barrier
formed by the angular momentum and a bottom formed by the gravity
on its right . (Right) The barrier top and the bottom merge at a special
scale (marked by a red point) and the BEC portion inside this scale
generates SMBH. }
\label{fig3}
\end{figure}

We now consider the attractive interaction ($g<0$) expecting that
this moderately promotes the SMBH formation by reducing the angular
momentum effect. A typical boson field with attractive force would
be the Axion, which is also a good candidate of DM\cite{Sikivie 2009}.
The Axion field would be initially uniform and forms BEC because of
its low temperature and small mass. We do not know the shape of the
developed over-density region, but let us simply assume it as the
quasi-isothermal distribution as a toy model, 
\begin{equation}
\rho\left(r\right)=\rho_{0}\left(1+\left(\frac{r}{r_{0}}\right)^{2}\right)^{-1},\label{eq:QITdensity}
\end{equation}
where the system is supposed to extend up to the radius $R$. We further
assume that the rigid rotation of the condensed system with a constant
angular velocity $\omega$. Then the angular momentum $J$ of the
region inside some radius $r$ is simply given by integrating this
rotation with the weight Eq.(\ref{eq:QITdensity}) to this radius.
We now consider the effective potential Eq.(\ref{eq:VeffgGA}) for
this system. Generally, this potential has a maximum and a minimum,
\begin{equation}
\frac{J^{2}\pm\sqrt{J^{4}-48\pi a_{s}GM^{6}m^{-3}\hbar^{2}}}{2GM^{3}},
\end{equation}
where $M$ is the mass inside the radius $r$ (Fig.\ref{fig3} Left).

If these maximum and the minimum coincide with each other, then the
barrier of the angular momentum disappears and the corresponding region
of BEC can collapse into BH. This radius is estimated as 
\begin{equation}
r_{hb}=\frac{2\sqrt{3\pi}a_{s}^{1/2}\hbar}{G^{1/2}m^{3/2}},
\end{equation}
which is fully given by microscopic constants and independent from
the DM mass and angular momentum $M,J$. Axion attractive force, despite
being very weak, just cancels the effective potential barrier formed
by the angular momentum at this scale (Fig.\ref{fig3} Right). The
part of the system inside of this radius $r_{hb}$ collapses to form
SMBH and the rest of the system would form DH surrounding the SMBH. 

A typical scales for the Axion and the galaxy 
\begin{equation}
m=10^{-5}\mathrm{eV},a_{s}=10^{-29}\mathrm{meter},r_{0}=1\mathrm{kpc},R=10\mathrm{kpc}
\end{equation}
yield the values
\begin{equation}
r_{hb}=108pc,M_{SMBH}/M_{DH}=4.2\times10^{-5},t_{BH}=6.3\times10^{4}year
\end{equation}
where $M_{SMBH}/M_{DH}$ is the mass ratio of the formed SMBH and
the surrounding DH, and $t_{BH}$ is the time scale of the SMBH formation. 

Although a SMBH is formed within a very short time scale in this scenario,
too many such SMBH would be formed near the center of the galaxy.
Because one BH is formed in the volume defined by the above scale
$r_{hb}$ everywhere in the galaxy. Then these dense SMBH soon collide
with each other to form larger SMBH. This coalescence process takes
place particularly in the central core regions of size $r_{0}$. Then
about $\left(r_{0}/r_{hb}\right)^{3}\approx10^{3}$ SMBH would coalesce
with each other to form much larger SMBH at the center of the galaxy. 

We can estimate the time scale of this whole process from the evaporation
time scale using the standard dynamical friction theory \cite{Chandrasekhar 1943}.
Assuming the Boltzmann distribution for an $N$-body self-gravitating
system, and that the portion of the positive energy particles in the
whole distribution, $\gamma\approx0.0074$, can escape from the cluster
to infinity, the time scale becomes 
\begin{equation}
\tau_{{\rm {d}}}^ {}=\frac{2}{{9\gamma}}\tau_{{\rm {r}}}\approx30.1\tau_{{\rm {r}}}^ {},\label{eq:t-dissipation}
\end{equation}
where the dynamical relaxation time scale is 
\begin{equation}
\tau_{{\rm {r}}}^ {}=\frac{N}{{12\sqrt{2}\ln N}}\tau_{{\rm {c}}},\label{eq:t-relaxiation}
\end{equation}
and the free-fall time scale is 

\begin{equation}
\tau_{{\rm {c}}}=\frac{R}{v}=\sqrt{\frac{{R_{}^{3}}}{{GM}}}.\label{eq:t-ff}
\end{equation}
 Then Eq.(\ref{eq:t-dissipation}) yields about $10^{8}$ years for
our case $N\approx10^{3}$, well within the observational constraints.

\begin{figure}
\includegraphics[width=7cm]{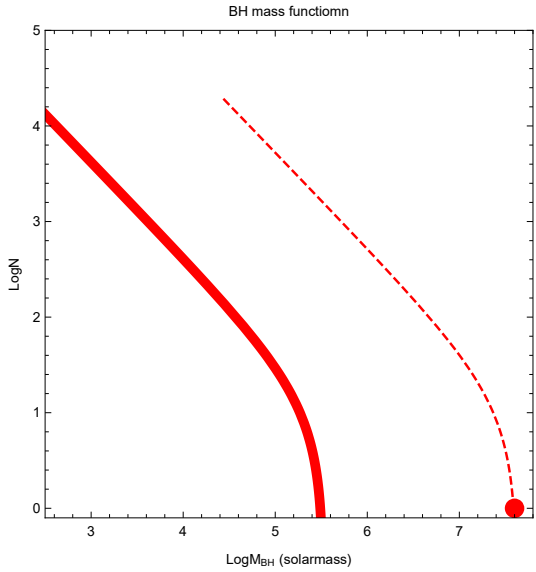}\caption{The mass distribution functions of BH within a single galaxy. The
broken line shows the case for $a_{s}=10^{-29}$meter, while the solid
line and the separate point show for $a_{s}=10^{-30.4}$meter. The
latter case represents a more realistic estimate for the SMBH formation
considering the coalescence at the central region by the dynamical
friction within the time scale $t_{d}=6.8\times10^{7}\mathrm{years}.$}
\label{fig4}
\end{figure}

Furthermore, in this Axion case, many smaller black holes, of mass
range $10^{2-5}M_{\odot}$, are formed as well in the outskirts of
the galaxy. The mass function of them has almost the power law distribution
and one dominant contribution from the SMBH (Fig.\ref{fig4}). This
is the mass distribution function of BH within a single galaxy and
should not be confused with the ordinary global mass function of BH
in the Universe usually expressed by the Press-Schechter function.
A galaxy turns out to be filled with plenty of black holes of various
masses in this case. 


\section{Conclusions and Discussions}

We have considered the scenario that the Bose-Einstein condensate
(BEC) dark matter (DM) collapses to form supermassive black holes
(SMBH) in the galactic center. This coherent wave easily collapses
to form SMBH. Actually, by using a simple Gaussian non-relativistic
approximation, we have obtain the collapsing dynamics of the isotropic
as well as anisotropic distributions of BEC. Thus SMBH is formed but
most of the BEC collapses leaving no dark halo (DH) in the surrounding
regions in these simple cases. On the other hand, if we introduce
the realistic amount of angular momentum, BEC does not collapse at
all to form BH and simply the whole DH just remains. Then, we have
introduced a small amount of attractive interaction of the condensed
bosons, such as Axion field. We have found that some portion near
the center of the BEC system collapses to form thousands of BHs which
eventually coalesce with each other by the dynamical relaxation and
form a single SMBH within the remaining huge amount of DH. A typical
mass ratio of them turns out to be $M_{SMBH}/M_{DH}\approx10^{-4}$.
This may yield a commonly observed SMBH-DH correlation. We have obtained
a tentative mass distribution function of BH within a single galaxy,
which should be compared with future observations. 

In order to complete this scenario of SMBH formation from BEC, the
following work is needed. Firstly, if we do not rely upon the Axion
model, we have to go back to the early Universe when the galaxy/halo
first obtained their angular momentum by the tidal torque mechanism.
Secondly, we have to study the repeated collapse and bounce of BEC
with diminishing amplitude, in relation to the general concentric
shell structure observed in many galaxies. Finally, we have to extend
our calculation based on the general relativity in order to clarify
the precise process and condition of the BH formation.

\end{document}